\documentclass[twocolumn,tighten]{aastex63}
\usepackage{amssymb, amsmath,framed}

\usepackage{bm}
\expandafter\ifx\csname package@font\endcsname\relax\else
 \expandafter\expandafter
 \expandafter\usepackage
 \expandafter\expandafter
 \expandafter{\csname package@font\endcsname}
\fi
\def\bq{\begin{equation}}
\def\eq{\end{equation}}
\def\bqy{\begin{eqnarray}}
\def\eqy{\end{eqnarray}}

\begin{document}
\title{\large{Role of Planetary Radius on Atmospheric Escape of Rocky Exoplanets}}

\correspondingauthor{Chuanfei Dong}
\email{dcfy@bu.edu}

\correspondingauthor{Manasvi Lingam} 
\email{mlingam@fit.edu}

\author[0009-0006-7877-1835]{Laura Chin}
\affiliation{Department of Astronomy, Boston University, Boston, MA 02215, USA}

\author[0000-0002-8990-094X]{Chuanfei Dong}
\affiliation{Department of Astronomy, Boston University, Boston, MA 02215, USA}

\author[0000-0002-2685-9417]{Manasvi Lingam}
\affiliation{Department of Aerospace, Physics and Space Sciences, Florida Institute of Technology, Melbourne, FL 32901, USA}
\affiliation{Department of Physics, The University of Texas at Austin, Austin, TX 78712, USA}

\begin{abstract}
Large-scale characterization of exoplanetary atmospheres is on the horizon, thereby making it possible in the future to extract their statistical properties. In this context, by using a well validated model in the solar system, we carry out three-dimensional magnetohydrodynamic simulations to compute nonthermal atmospheric ion escape rates of unmagnetized rocky exoplanets as a function of their radius based on fixed stellar radiation and wind conditions. We find that the atmospheric escape rate is, unexpectedly and strikingly, a nonmonotonic function of the planetary radius $R$ and that it evinces a maximum at $R \sim 0.7\,R_\oplus$. This novel nonmonotonic behavior may arise from an intricate tradeoff between the cross-sectional area of a planet (which increases with size, boosting escape rates) and its associated escape velocity (which also increases with size, but diminishes escape rates). Our results could guide forthcoming observations because worlds with certain values of $R$ (such as $R \sim 0.7\,R_\oplus$) might exhibit comparatively higher escape rates when all other factors are constant. \\
\end{abstract}

\section{Introduction}\label{SecIntro}
The number of confirmed exoplanets has grown explosively in the past decade \citep{WF15,MP18,ZD21}, with the current total exceeding $5000.$\footnote{\url{https://exoplanetarchive.ipac.caltech.edu/}} Current and forthcoming surveys are anticipated to raise the number of detected and characterized exoplanets by a substantive degree. As a result of the relatively large sample size, it has become feasible to extract exoplanetary statistics and trends.

One of the most noteworthy among these discoveries is the sparsity of short-period planets with radii $\sim 2\,R_\oplus$, often termed a radius ``gap'' or ``valley'' or ``desert'' \citep{LF14,FPH17,VAL18,MCG19,MKL19,BHG,CM20,BRO21}, with smaller worlds typically referred to as super-Earths and larger planets known as sub-Neptunes. The mechanisms propounded for explaining this bimodal distribution include extreme ultraviolet (EUV) and X-ray photoevaporation \citep{CR16,OW17,FP18,M20}, and core-powered mass loss \citep{GSS,GS19,GS20}, and synergies thereof \citep{OS23}, among other candidates. Differentiating between various processes that could sculpt this valley may require large-scale surveys that sample $\gtrsim 5000$ stellar systems \citep{RGO21}.

In a similar vein, statistics of terrestrial planets -- roughly interpreted to possess radii $< 2\,R_\oplus$ -- situated within or close to the habitable zone (HZ) are likely to become increasingly relevant in the coming decades; the limits of the HZ were computed in \citet{KWR93} and \citet{KRK13,KRS14}. Rocky planets in the HZ can host liquid water on their surfaces, and such worlds are likely to be detected and characterized with increasing frequency by future telescopes \citep{FAD18,NM19,WK22}. The atmospheres of these terrestrial exoplanets may yield a wealth of data about their physical, chemical, and geological properties \citep{SD10,NM19,XZ20,TJCI2023}, as well as potentially even biological features \citep{FAD18}.

Given the centrality of exoplanetary atmospheres, especially those of rocky exoplanets proximal to the HZ as delineated above, it is of crucial importance to understand what \emph{statistical} trends, if any, may be discernible by future surveys. This era has already commenced, with JWST having obtained the transmission spectrum or detected the thermal emission of some warm rocky exoplanets \citep{LFM23,GBD23,ZKD23}. From the perspective of atmospheric loss -- which could sculpt the atmospheres of terrestrial planets \citep{FT15,JO19} -- the significance of nonthermal atmospheric ion escape processes driven by the solar (or stellar) wind for this class of planets is well-established in our solar system \citep{LKC08,HL13,BBM,DLM18,PFR20,RB21,LDS} and is plausible for exoplanets (as reviewed in \citealt{ZC17,LL19,ML21,ABC20,GAB20}).

In this work, we carry out a systematic investigation of ``nonthermal'' atmospheric ion escape -- in which the speeds of escaping particles are essentially decoupled from the exobase temperature, and often entail the presence of ions in electromagnetic fields \citep{FT15,LDO20} -- as a function of radius for unmagnetized rocky exoplanets with atmospheric composition akin to Venus. The latter world is recognized as a valuable benchmark for exoplanetary science \citep{KAC19,SRK22}, especially insofar as planets close to the inner edge of the HZ are concerned \citep{OKL23}. The outline of the Letter is as follows. We describe the multi-species MHD model and the setup in Section \ref{SecMod}. Next, we present the results and analyze the ensuing implications in Section \ref{SecRes}. We round off by summarizing our findings in Section \ref{SecConc}.

\section{Methods and model setup}\label{SecMod}
We leverage the 3-D Block-Adaptive-Tree Solarwind Roe-type Upwind Scheme (BATS-R-US) multispecies magnetohydrodynamic (MS-MHD) model that has been successfully implemented to model Venus-like exoplanets \citep{DLMC,DJL18,DHL19,DJL20} for simulating atmospheric ion escape from exoplanets spanning a range of radii. The MS-MHD model includes four ion species H$^+$, O$^+$, O$_2^+$, CO$_2^+$, and the associated ionospheric photochemistry such as photoionization, charge exchange, and electron recombination. For unmagnetized planets, the BATS-R-US MS-MHD model investigates the interactions of magnetized stellar winds with planetary upper atmospheres, and properly accounts for atmospheric escape mechanisms mediated by the stellar wind via mass loading.

In addition to assuming Venusian atmospheric composition (i.e., CO$_2$-dominated atmosphere) due to the reason outlined in the last paragraph of Section \ref{SecIntro}, the model accepts several input parameters to specify planetary and stellar properties, including but not limited to planetary radius and mass, atmospheric profile, as well as stellar winds and radiation. We have chosen to work with the Planet-Star-Orbital (PSO) coordinates, where the X-axis points from the planet toward the star, the Z-axis is perpendicular to the planetary orbital plane, and the Y-axis completes the right-hand system.

{\centering
\begin{table}
\begin{tabular}{ c  c  c }
\hline
    Radius ($R_{V}$) & Radius (m) & Mass (kg) \\
\hline
    0.50 & 3026E+3 & 0.374E+24 \\
    0.75 & 4539E+3 & 1.678E+24 \\
    1.00 & 6052E+3 & 4.865E+24 \\
    1.25 & 7565E+3 & 11.11E+24 \\
    1.50 & 9078E+3 & 21.81E+24 \\
    1.75 & 10591E+3 & 38.58E+24 \\
    2.00 & 12104E+3 & 63.23E+24 \\
    2.25 & 13617E+3 & 97.76E+24 \\
\hline
\end{tabular}
\caption{Model exoplanet parameters for rocky exoplanets. Given that $R_{V} \approx R_\oplus$, the sizes of these putative planets span the range of around half to twice the size of Earth.}
\label{table:ModelParams}
\end{table}}

We modeled eight exoplanets ranging in size from $0.50$ to $2.25$ Venus radii ($R_V$), as illustrated in Table \ref{table:ModelParams}. The lower bound is approximately the size of Mars, which was capable of retaining an atmosphere at early epochs \citep{EAA16}. This upper bound is compatible with empirical and theoretical constraints on the size of the largest worlds with rocky composition \citep{OBH20,PV20}. Each model planet has a surface pressure of $1$ bar; accordingly, larger planets possess more massive atmospheres, which is consistent with intuition. While we could opt to vary the surface pressure, the previous work \citep{DLMC} has demonstrated that the nonthermal atmospheric ion escape rates do not significantly rely on planetary surface pressure; in other words, the choice of surface pressure is not expected to conspicuously affect our calculations on the atmospheric ion escape rates.

For each model planet, we modify the neutral atmospheric scale height ($H$) by varying the gravitational acceleration ($g$) since the former is defined as
\begin{equation}
    H = \frac{kT}{m g},\label{eq1}
\end{equation}
where $T$ is the temperature, $k$ is the Boltzmann constant, and $m$ is the mass of the atmospheric species. Assuming that the planetary radius $R$ is known, the mass of the planet $M$ is derived from the simplified mass-radius relation for rocky planets (\citealt{VOS06,ZSJ16,ZJS19}; see also \citealt{CK17}):
\begin{equation}
    \frac{R}{R_{V}} \approx \left(\frac{M}{M_{V}}\right)^{1/3.7},
\end{equation}
where $M_V$ and $R_V$ are the mass and radius of Venus, respectively. Given $M$ and $R$, it is found that $g$ obeys
\begin{equation}
    g = \frac{GM}{R^{2}} \propto R^{1.7}.\label{eq3}
\end{equation}
In our study, the model exoplanets are assumed to be unmagnetized. This assumption not only matches the current state of Venus, but also might be characteristic of $\sim 50\%$ of all rocky planets in the HZ as per some estimates \citep{MLI19}. However, future work is needed to assess the impact of a planetary magnetic field on nonthermal ion escape \citep{DLMC,GMN18,EJM19,LM19,DJL20}, as those effects are not yet fully understood.

Next, a crucial set of parameters that needs to be specified is that of the stellar wind. In this context, it should be borne in mind that M-dwarfs not only constitute the majority of stars in the Milky Way, but also their (rocky) planets represent promising targets for atmospheric characterization \citep{SBJ16,FAD18,ML21}. Although we do not simulate the conditions experienced by M-dwarf exoplanets (which are diverse) as such, we instead choose to work with ancient solar wind conditions (near Earth) at $\sim 4$ Ga consistent with high mass-loss rate of the young Sun \citep{WMR21} -- with solar wind density n$_{sw}$ = 136.7 cm$^{-3}$, solar wind velocity \textbf{v}$_{sw}$ = (-910, 0, 0) km/s, and interplanetary magnetic field \textbf{B}$_{sw}$ = (-15.6, 30.2, 0) nT -- and a stellar EUV flux that is $12$ times that of modern Sun \citep{BLK10,DHL17} for our model exoplanets. To an extent, these enhanced stellar wind parameters and EUV flux are reminiscent of those encountered by M-dwarf exoplanets in the HZ \citep{FFL13,FLY16,ABC20}.

\section{Results and Discussion}\label{SecRes}
In this section, we cover the salient results and provide a plausible explanation for the discerned trends.

\subsection{Results}\label{SSecRes}

\begin{figure}[h]
  \centering
  \includegraphics[width=0.48\textwidth]{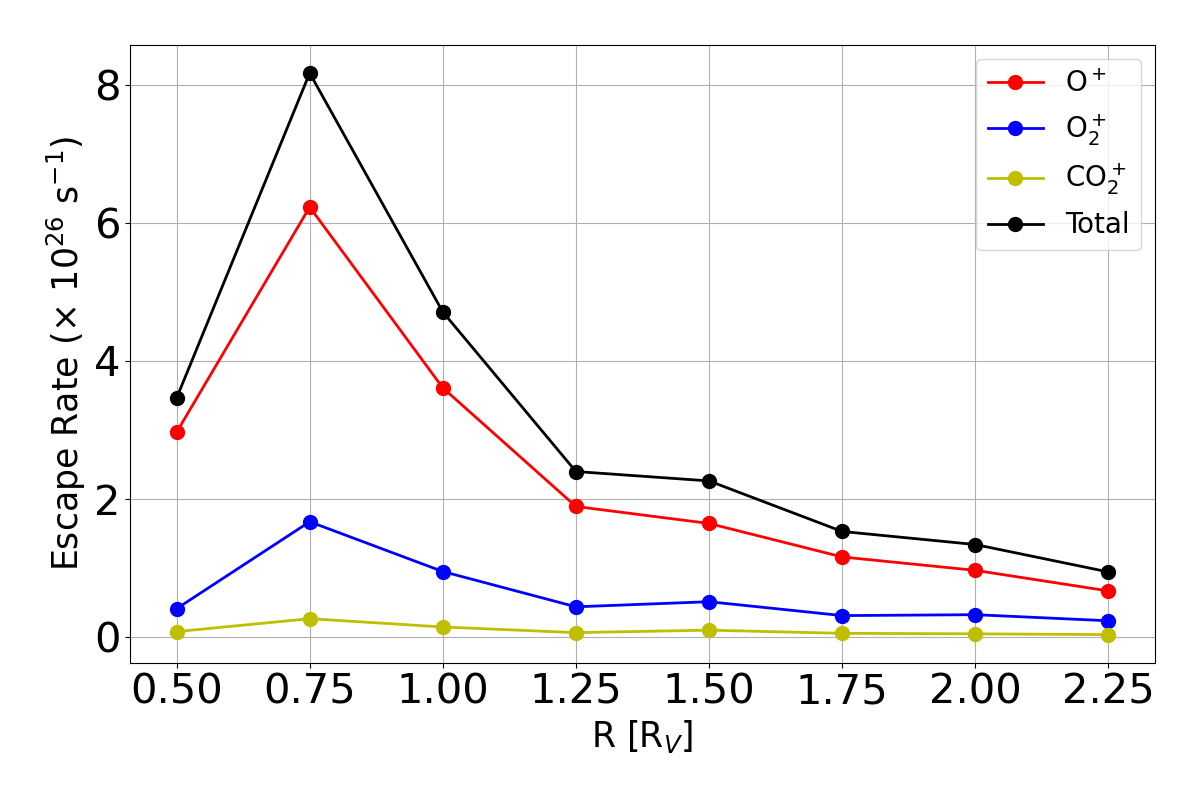}
\caption{The calculated atmospheric ion escape rates as a function of planetary radius, which manifest a nonmonotonic trend, peaking at $R=0.75 R_{V}$.}
\label{fig:ResultsFourCurves}
\end{figure}

For the model exoplanets listed in Table \ref{table:ModelParams}, we deployed the BATS-R-US MS-MHD model and computed the associated atmospheric ion escape rates; the latter were calculated based on the stellar radiation and wind conditions motivated toward the end of Section \ref{SecMod}. The atmospheric ion escape rates are reported in Figure \ref{fig:ResultsFourCurves} and Table \ref{table:AncientIonEscapeRates} accordingly illustrates the relationship between planetary radius and atmospheric ion escape rate. 

\begin{figure*}
\centering
\includegraphics[width=1.00\textwidth]{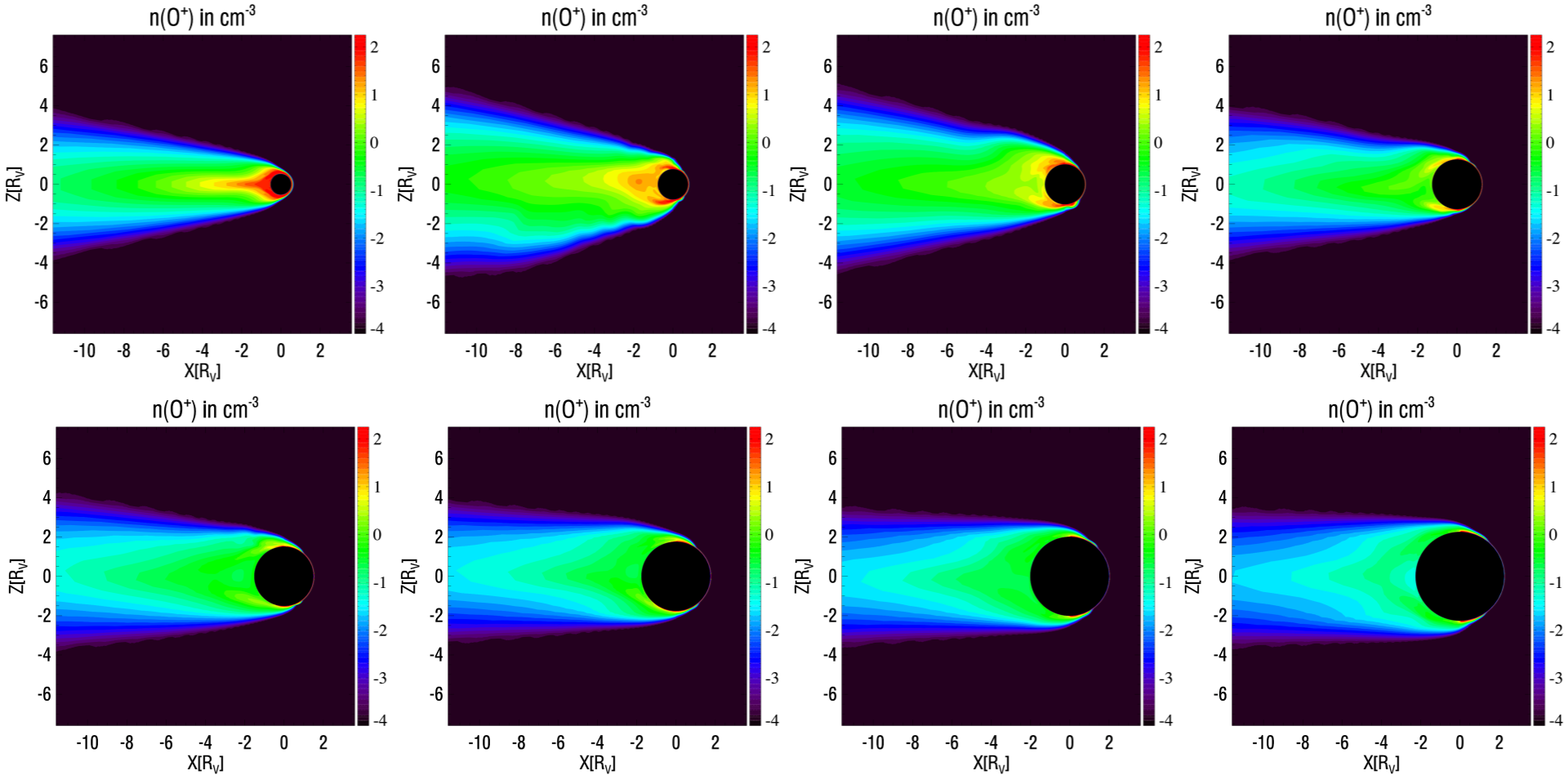}
\caption{Logarithmic scale contour plots of the escaping O$^{+}$ ion density (units of cm$^{-3}$) in the meridional plane for cases with different planetary radii subjected to the same ancient solar conditions. The second plot from the left in the top row ($R=0.75$ $R_{V}$) has the highest atmospheric ion escape rate.}
\label{fig:ContourPlots}
\end{figure*}

On inspecting Figure \ref{fig:ResultsFourCurves}, it is apparent that the relatively light ion species O$^+$ has the highest ion escape rate and the relatively heavy species CO$_{2}^+$ has the lowest ion escape rate. This trend is potentially explainable by the fact that lighter ions like O$^+$ are typically more abundant at higher altitudes (as reviewed in \citealt{MWD18}), in the absence of strong (e.g., turbulent) mixing, and are consequently more vulnerable to stellar wind erosion. Hence, given that the stellar wind could strip a relatively higher concentration of lighter ions, it cannot also strip heavier ions at lower altitudes with the same efficiency due to the constraint imposed by the conservation of momentum and energy.

In examining Figure \ref{fig:ResultsFourCurves}, we identify an intriguing nonmonotonic feature in the trend of the ion escape rate as a function of the radius (of rocky planets). For the chosen stellar wind conditions, which are loosely reminiscent of those parameters that may be experienced by temperate M-dwarf exoplanets, total ion escape rate peaks at $R=0.75 R_{V}$. The maximum total ion escape rate at $R=0.75 R_{V}$ is about eight times higher than the minimum escape rate at $R=2.25 R_{V}$ (see Table \ref{table:AncientIonEscapeRates}). The putative rationale for the former feature and its implications are discussed shortly hereafter in Section \ref{sec:Analysis}.

Figure \ref{fig:ContourPlots} depicts O$^+$ ion density contour plots for all model exoplanets subjected to the same stellar radiation and wind conditions. The black sphere in the middle represents the planetary body. The planetary radius increases from $R=0.50 R_{V}$ (top left) to $R=2.25 R_{V}$ (bottom right) with $\Delta R=0.25 R_{V}$. It is clear that the escaping $O^+$ ion density around the planetary body with $R=0.5 R_{V}$ has the highest value, but the escape rate peaks at $R=0.75 R_{V}$ (see Table \ref{table:AncientIonEscapeRates}). It is noteworthy that the escape rates in Figure \ref{fig:ResultsFourCurves} and Table \ref{table:AncientIonEscapeRates} are calculated based on the integration of escape ion flux over a spherical surface around the planet, thus the size of the planet is also important. When planetary radius exceeds $R=1.25 R_{V}$, one can barely observe the red contours around the planet due to the increasing surface gravity.   

{\centering
\begin{table}
\begin{tabular}{c  c  c  c  c}
\hline
    R ($R_{V}$) & $O^+$ & $O_{2}^+$ & $CO_{2}^+$ & Total\\
\hline
    0.50 & 2.967 & 0.413 & 0.081 & 3.462 \\
    0.75 & 6.229 & 1.672 & 0.268 & 8.169 \\
    1.00 & 3.611 & 0.951 & 0.148 & 4.710 \\
    1.25 & 1.893 & 0.440 & 0.066 & 2.399 \\
    1.50 & 1.647 & 0.514 & 0.102 & 2.263 \\
    1.75 & 1.162 & 0.313 & 0.056 & 1.531 \\
    2.00 & 0.968 & 0.326 & 0.049 & 1.342 \\
    2.25 & 0.670 & 0.238 & 0.037 & 0.945 \\
\hline
\end{tabular}
\caption{The calculated atmospheric ion escape rates (of various ion species) in units of $10^{26} s^{-1}$ as a function of the planetary radius.}
\label{table:AncientIonEscapeRates}
\end{table}}

\subsection{Analysis}\label{sec:Analysis}
One of the most striking results manifested is the nonmonotonic variation of the atmospheric escape rate with the planetary radius $R$. We begin by furnishing a possible qualitative explanation for this behavior, and then elaborate further on this theme.

As we are studying atmospheric escape, there are two competing factors at play. Depending on which of the duo is more dominant, the escape rate may be altered accordingly. On the one hand, the cross-sectional area of the planet that interacts with the stellar wind is of importance. In the case of weakly magnetized and wholly unmagnetized planets, this area is roughly proportional to $R^2$ \citep[cf.][]{SSC16}, although for magnetized planets the expression is more complex \citep{BT18,LM19}. Therefore, as the radius is elevated, the cross-sectional area increases, and so does the escape rate because the latter scales with the area. Likewise, it is important to recognize that the area of the atmosphere -- the source from which the particles escape the planet -- is also proportional to $R^2$.

On the other hand, as the radius is increased, the escape velocity of the planet ($v_e$) is also enhanced:
\begin{equation}\label{EscVel}
    v_e = \sqrt{\frac{2 G M}{R}} \propto R^{1.35},
\end{equation}
where we have used the mass-radius relationship $M \propto R^{3.7}$ for rocky planets \citep{ZSJ16}. On account of the higher escape velocity (at larger $R$), it would be harder for particles to escape the planet's gravitational well. In such a scenario, the atmospheric escape rate is presumed to steeply decline with the radius, which runs counter to the trend in the preceding paragraph. Hence, when these two aspects are duly taken into consideration, the nonmonotonic behavior of the atmospheric escape rate as a function of the planetary radius exhibited by our simulations might be explainable.

It is possible to build on this theme. For an unmagnetized planet with a fixed radius, its cross-sectional area $\pi R^2$ determines the amount of stellar radiation and wind energy incident on the planet. The larger the cross-sectional area is, the more energy the planet ought to receive. The stellar radiation energy can lead to photoionization (the main source of the ionized gases) and heating of the upper atmosphere, and subsequently, the stellar wind could strip those ionized particles from the planet. Given that it is relatively easy for ionized gas to escape from a small unmagnetized planet (with lower gravity), the atmospheric ion escape may become source limited, i.e., almost all ions supplied to the region energized by the stellar wind can gain sufficient energy to escape, and therefore, the source of the ions (i.e., controlled by the ion production rate) could limit the supply and thence the total ion escape flux, rather than the amount of available energy. Therefore, this scenario enables a trend whereby larger unmagnetized planets exhibit higher atmospheric ion escape rates due to bigger reservoirs of ionized gas in their upper atmospheres. 

On the other hand, as the planet's size is further increased, the escape velocity is also enhanced, as revealed by (\ref{EscVel}). Therefore, beyond a certain threshold, it would no longer be easy for atmospheric ions to attain the requisite energy for escaping the planet; to put it differently, the escape rate might transition from a source-limited regime to an energy-limited regime. A brief segue is warranted at this stage. The energy-limited regime is encountered often in hydrodynamic escape \citep[e.g.,][]{WDW81,EKL07,OJ12,OA16,SSC16,KFE18,KFKL,LLC21}. However, in contrast to hydrodynamic escape on worlds with H/He atmospheres, it is worth recognizing that the adopted EUV flux alone is unlikely to drive atmospheric mass loss on Venus-like planets for two major reasons: (1) the atmosphere is chiefly composed of heavy chemical species; and (2) CO$_2$ 15-$\mu$m cooling is very efficient in the upper atmosphere of Venus-like planets \citep{Bougher1994}. In this paper, the effects of EUV flux are, instead, to produce photoionized gas and heat the upper atmosphere (through photoionization and photoelectron heating); subsequently, the stellar wind can strip the ionized gas away \citep{LDO20}.

Circling back to the above theme, the bottleneck on larger rocky planets is now envisioned to be imposed by the escape velocity, which is a monotonically increasing function of $R$ -- refer to (\ref{EscVel}). Hence, we would expect the escape rate to strongly decline with an increase in the radius across this range, as confirmed by Figure \ref{fig:ResultsFourCurves} and Table \ref{table:AncientIonEscapeRates}. On combining these predictions, the escape rate ought to first increase and then decrease with the radius, which is consistent with Figure \ref{fig:ResultsFourCurves} and Table \ref{table:AncientIonEscapeRates}.

To sum up, we have argued that we could witness a change in the regime, going from a limit set by source particles to one enforced by energy. Quite strikingly, such a shift has been empirically documented in the solar system, where nonthermal ion escape on Mars is source-limited, whereas the equivalent on Venus is energy-limited \citep{PFR20,RB21}; note that both these planets are weakly magnetized, allowing for a comparison of similar worlds. This data gives credence to the preceding paragraphs, although we caution that we have offered a simplified analysis, which was necessitated by the complexity of nonthermal escape mechanisms \citep{CRF17,LL19,ABC20,GAB20}.

In closing, we comment on the ramifications of our modeling. If the nonthermal ion escape rates do indeed peak at $R \sim 0.7\,R_\oplus$, this datum could have crucial consequences for future observations. Given that the computed escape rate is boosted by a factor of about $2$ as we transition from $R \sim 0.5\,R_\oplus$ to $R \sim 0.7\,R_\oplus$ and declines thereafter by nearly the same amount when we move to $R > R_\oplus$, it is possible that temperate rocky planets with $R \sim 0.7\,R_\oplus$ might harbor relatively thinner atmospheres if all other variables (e.g., outgassing, atmospheric mass) are statistically similar across the range of planetary radii.\footnote{The exact value of this peak may, however, change based on the stellar parameters (e.g., EUV and stellar wind), and will need to evaluated on a case-by-case basis.} If future telescopes such as ARIEL \citep{TDE} characterize a sufficiently large sample of terrestrial exoplanets, it may be feasible to test some aspects of our hypothesis empirically.

\section{Conclusions}\label{SecConc}
Building on the ongoing characterization of exoplanetary atmospheres by JWST \citep{GMA23} -- which has led to several interesting discoveries \citep[e.g.,][]{LFM23,GBD23,ZKD23} -- and observations by forthcoming ground-based extremely large telescopes, a wealth of data will become available, making it feasible to discern statistical patterns in exoplanetary atmospheres in the future. Nonthermal ion escape facilitated by stellar EUV radiation and driven by the stellar wind represents one of the key regulators of the mass and composition of terrestrial exoplanetary atmospheres \citep{BBM,ZC17,ABC20,GAB20}. Motivated by these facts, we utilized a thoroughly validated multispecies MHD model to investigate the total ion escape rate as a function of the planetary radius $R$.

After performing the simulations, we unearthed a novel trend for rocky planets with CO$_2$-dominated atmospheres: the escape rate is a nonmonotonic function of $R$, as clearly depicted in Figure \ref{fig:ResultsFourCurves}, with a peak at $R \sim 0.7\,R_\oplus$. It was found that this characteristic is manifested for intense stellar wind and radiation conditions, which may be associated with M-dwarfs and young Sun-like stars. This nonmonotonic feature runs counter to the naive expectation that smaller rocky planets (e.g., akin to Mars) would automatically exhibit higher escape rates because of their weaker gravity. 

The nonmonotonic behavior may arise from a tradeoff between the (cross-sectional) area and the escape velocity of a planet; both of these variables increase for larger planets. On the one hand, when the aforementioned area is boosted, this will enhance the atmospheric ion escape rate since the planet would intercept more of the stellar radiation and wind (due to its greater area), thereupon amplifying the escape rate on account of the enhanced ionized reservoir in the planetary upper atmosphere. On the other hand, when the escape velocity is boosted, this factor will suppress the escape rate because particles require more energy to escape the gravitational well. From a more detailed perspective, we suggested in Section \ref{sec:Analysis} that nonthermal atmospheric ion escape may shift from a source-limited regime to an energy-limited one, which could partially explain the simulated results.

If the striking trend displayed by our modeling is correct, it may have tangible observational consequences for future surveys of terrestrial exoplanets. Provided that all other factors are held equal, a higher atmospheric ion escape rate could translate to relatively thinner atmospheres. Hence, if future surveys either discover thinner atmospheres -- or, even better, obtain direct evidence of higher nonthermal ion escape rates -- at a certain value of $R$ (conceivably $R \sim 0.7\,R_\oplus$), such data would serve to corroborate the predictions of our numerical simulations. Moreover, such nonmonotonic behavior would deepen our understanding of how nonthermal atmospheric escape can sculpt the properties of exoplanetary atmospheres of rocky worlds.

\acknowledgments
This work was supported by NASA grants 80NSSC23K1115 and 80NSSC23K0911. Resources supporting this work were provided by the NASA High-End Computing (HEC) Program through the NASA Advanced Supercomputing (NAS) Division at Ames Research Center. The Space Weather Modeling Framework that comprises the BATS-R-US code used in this study is publicly available at \url{https://github.com/MSTEM-QUDA/BATSRUS}.

\bibliographystyle{aasjournal}

\begin{thebibliography}{}
\expandafter\ifx\csname natexlab\endcsname\relax\def\natexlab#1{#1}\fi
\providecommand{\url}[1]{\href{#1}{#1}}
\providecommand{\dodoi}[1]{doi:~\href{http://doi.org/#1}{\nolinkurl{#1}}}
\providecommand{\doeprint}[1]{\href{http://ascl.net/#1}{\nolinkurl{http://ascl.net/#1}}}
\providecommand{\doarXiv}[1]{\href{https://arxiv.org/abs/#1}{\nolinkurl{https://arxiv.org/abs/#1}}}

\bibitem[{{Airapetian} {et~al.}(2020){Airapetian}, {Barnes}, {Cohen},
  {Collinson}, {Danchi}, {Dong}, {Del Genio}, {France}, {Garcia-Sage},
  {Glocer}, {Gopalswamy}, {Grenfell}, {Gronoff}, {G{\"u}del}, {Herbst},
  {Henning}, {Jackman}, {Jin}, {Johnstone}, {Kaltenegger}, {Kay}, {Kobayashi},
  {Kuang}, {Li}, {Lynch}, {L{\"u}ftinger}, {Luhmann}, {Maehara}, {Mlynczak},
  {Notsu}, {Osten}, {Ramirez}, {Rugheimer}, {Scheucher}, {Schlieder},
  {Shibata}, {Sousa-Silva}, {Stamenkovi{\'c}}, {Strangeway}, {Usmanov},
  {Vergados}, {Verkhoglyadova}, {Vidotto}, {Voytek}, {Way}, {Zank}, \&
  {Yamashiki}}]{ABC20}
{Airapetian}, V.~S., {Barnes}, R., {Cohen}, O., {et~al.} 2020, Int. J.
  Astrobiol., 19, 136, \dodoi{10.1017/S1473550419000132}

\bibitem[{{Bean} {et~al.}(2021){Bean}, {Raymond}, \& {Owen}}]{BRO21}
{Bean}, J.~L., {Raymond}, S.~N., \& {Owen}, J.~E. 2021, J. Geophys. Res.
  Planets, 126, e06639, \dodoi{10.1029/2020JE006639}

\bibitem[{{Berger} {et~al.}(2020){Berger}, {Huber}, {Gaidos}, {van Saders}, \&
  {Weiss}}]{BHG}
{Berger}, T.~A., {Huber}, D., {Gaidos}, E., {van Saders}, J.~L., \& {Weiss},
  L.~M. 2020, Astron. J., 160, 108, \dodoi{10.3847/1538-3881/aba18a}

\bibitem[{{Blackman} \& {Tarduno}(2018)}]{BT18}
{Blackman}, E.~G., \& {Tarduno}, J.~A. 2018, Mon. Not. R. Astron. Soc., 481,
  5146, \dodoi{10.1093/mnras/sty2640}

\bibitem[{{Boesswetter} {et~al.}(2010){Boesswetter}, {Lammer}, {Kulikov},
  {Motschmann}, \& {Simon}}]{BLK10}
{Boesswetter}, A., {Lammer}, H., {Kulikov}, Y., {Motschmann}, U., \& {Simon},
  S. 2010, Planet. Space Sci., 58, 2031, \dodoi{10.1016/j.pss.2010.10.003}

\bibitem[{{Bougher} {et~al.}(1994){Bougher}, {Hunten}, \&
  {Roble}}]{Bougher1994}
{Bougher}, S.~W., {Hunten}, D.~M., \& {Roble}, R.~G. 1994, J. Geophys. Res.,
  99, 14609, \dodoi{10.1029/94JE01088}

\bibitem[{{Brain} {et~al.}(2016){Brain}, {Bagenal}, {Ma}, {Nilsson}, \&
  {Stenberg Wieser}}]{BBM}
{Brain}, D.~A., {Bagenal}, F., {Ma}, Y.~J., {Nilsson}, H., \& {Stenberg
  Wieser}, G. 2016, J. Geophys. Res. Planets, 121, 2364,
  \dodoi{10.1002/2016JE005162}

\bibitem[{{Chen} \& {Rogers}(2016)}]{CR16}
{Chen}, H., \& {Rogers}, L.~A. 2016, Astrophys. J., 831, 180,
  \dodoi{10.3847/0004-637X/831/2/180}

\bibitem[{{Chen} \& {Kipping}(2017)}]{CK17}
{Chen}, J., \& {Kipping}, D. 2017, Astrophys. J., 834, 17,
  \dodoi{10.3847/1538-4357/834/1/17}

\bibitem[{{Cloutier} \& {Menou}(2020)}]{CM20}
{Cloutier}, R., \& {Menou}, K. 2020, Astron. J., 159, 211,
  \dodoi{10.3847/1538-3881/ab8237}

\bibitem[{{Cravens} {et~al.}(2017){Cravens}, {Rahmati}, {Fox}, {Lillis},
  {Bougher}, {Luhmann}, {Sakai}, {Deighan}, {Lee}, {Combi}, \&
  {Jakosky}}]{CRF17}
{Cravens}, T.~E., {Rahmati}, A., {Fox}, J.~L., {et~al.} 2017, J. Geophys. Res.
  Space Phys., 122, 1102, \dodoi{10.1002/2016JA023461}

\bibitem[{{Dong} {et~al.}(2019){Dong}, {Huang}, \& {Lingam}}]{DHL19}
{Dong}, C., {Huang}, Z., \& {Lingam}, M. 2019, Astrophys. J. Lett., 882, L16,
  \dodoi{10.3847/2041-8213/ab372c}

\bibitem[{{Dong} {et~al.}(2017{\natexlab{a}}){Dong}, {Huang}, {Lingam},
  {T{\'o}th}, {Gombosi}, \& {Bhattacharjee}}]{DHL17}
{Dong}, C., {Huang}, Z., {Lingam}, M., {et~al.} 2017{\natexlab{a}}, Astrophys.
  J. Lett., 847, L4, \dodoi{10.3847/2041-8213/aa8a60}

\bibitem[{{Dong} {et~al.}(2020){Dong}, {Jin}, \& {Lingam}}]{DJL20}
{Dong}, C., {Jin}, M., \& {Lingam}, M. 2020, Astrophys. J. Lett., 896, L24,
  \dodoi{10.3847/2041-8213/ab982f}

\bibitem[{{Dong} {et~al.}(2018{\natexlab{a}}){Dong}, {Jin}, {Lingam},
  {Airapetian}, {Ma}, \& {van der Holst}}]{DJL18}
{Dong}, C., {Jin}, M., {Lingam}, M., {et~al.} 2018{\natexlab{a}}, Proc. Natl.
  Acad. Sci., 115, 260, \dodoi{10.1073/pnas.1708010115}

\bibitem[{{Dong} {et~al.}(2017{\natexlab{b}}){Dong}, {Lingam}, {Ma}, \&
  {Cohen}}]{DLMC}
{Dong}, C., {Lingam}, M., {Ma}, Y., \& {Cohen}, O. 2017{\natexlab{b}},
  Astrophys. J. Lett., 837, L26, \dodoi{10.3847/2041-8213/aa6438}

\bibitem[{{Dong} {et~al.}(2018{\natexlab{b}}){Dong}, {Lee}, {Ma}, {Lingam},
  {Bougher}, {Luhmann}, {Curry}, {Toth}, {Nagy}, {Tenishev}, {Fang},
  {Mitchell}, {Brain}, \& {Jakosky}}]{DLM18}
{Dong}, C., {Lee}, Y., {Ma}, Y., {et~al.} 2018{\natexlab{b}}, Astrophys. J.
  Lett., 859, L14, \dodoi{10.3847/2041-8213/aac489}

\bibitem[{{Egan} {et~al.}(2019){Egan}, {Jarvinen}, {Ma}, \& {Brain}}]{EJM19}
{Egan}, H., {Jarvinen}, R., {Ma}, Y., \& {Brain}, D. 2019, Mon. Not. R. Astron.
  Soc., 488, 2108, \dodoi{10.1093/mnras/stz1819}

\bibitem[{{Ehlmann} {et~al.}(2016){Ehlmann}, {Anderson}, {Andrews-Hanna},
  {Catling}, {Christensen}, {Cohen}, {Dressing}, {Edwards}, {Elkins-Tanton},
  {Farley}, {Fassett}, {Fischer}, {Fraeman}, {Golombek}, {Hamilton}, {Hayes},
  {Herd}, {Horgan}, {Hu}, {Jakosky}, {Johnson}, {Kasting}, {Kerber}, {Kinch},
  {Kite}, {Knutson}, {Lunine}, {Mahaffy}, {Mangold}, {McCubbin}, {Mustard},
  {Niles}, {Quantin-Nataf}, {Rice}, {Stack}, {Stevenson}, {Stewart}, {Toplis},
  {Usui}, {Weiss}, {Werner}, {Wordsworth}, {Wray}, {Yingst}, {Yung}, \&
  {Zahnle}}]{EAA16}
{Ehlmann}, B.~L., {Anderson}, F.~S., {Andrews-Hanna}, J., {et~al.} 2016, J.
  Geophys. Res. Planets, 121, 1927, \dodoi{10.1002/2016JE005134}

\bibitem[{{Erkaev} {et~al.}(2007){Erkaev}, {Kulikov}, {Lammer}, {Selsis},
  {Langmayr}, {Jaritz}, \& {Biernat}}]{EKL07}
{Erkaev}, N.~V., {Kulikov}, Y.~N., {Lammer}, H., {et~al.} 2007, Astron.
  Astrophys., 472, 329, \dodoi{10.1051/0004-6361:20066929}

\bibitem[{{France} {et~al.}(2013){France}, {Froning}, {Linsky}, {Roberge},
  {Stocke}, {Tian}, {Bushinsky}, {D{\'e}sert}, {Mauas}, {Vieytes}, \&
  {Walkowicz}}]{FFL13}
{France}, K., {Froning}, C.~S., {Linsky}, J.~L., {et~al.} 2013, Astrophys. J.,
  763, 149, \dodoi{10.1088/0004-637X/763/2/149}

\bibitem[{{France} {et~al.}(2016){France}, {Loyd}, {Youngblood}, {Brown},
  {Schneider}, {Hawley}, {Froning}, {Linsky}, {Roberge}, {Buccino},
  {Davenport}, {Fontenla}, {Kaltenegger}, {Kowalski}, {Mauas}, {Miguel},
  {Redfield}, {Rugheimer}, {Tian}, {Vieytes}, {Walkowicz}, \&
  {Weisenburger}}]{FLY16}
{France}, K., {Loyd}, R.~O.~P., {Youngblood}, A., {et~al.} 2016, Astrophys. J.,
  820, 89, \dodoi{10.3847/0004-637X/820/2/89}

\bibitem[{{Fujii} {et~al.}(2018){Fujii}, {Angerhausen}, {Deitrick},
  {Domagal-Goldman}, {Grenfell}, {Hori}, {Kane}, {Pall{\'e}}, {Rauer},
  {Siegler}, {Stapelfeldt}, \& {Stevenson}}]{FAD18}
{Fujii}, Y., {Angerhausen}, D., {Deitrick}, R., {et~al.} 2018, Astrobiology,
  18, 739, \dodoi{10.1089/ast.2017.1733}

\bibitem[{{Fulton} \& {Petigura}(2018)}]{FP18}
{Fulton}, B.~J., \& {Petigura}, E.~A. 2018, Astron. J., 156, 264,
  \dodoi{10.3847/1538-3881/aae828}

\bibitem[{{Fulton} {et~al.}(2017){Fulton}, {Petigura}, {Howard}, {Isaacson},
  {Marcy}, {Cargile}, {Hebb}, {Weiss}, {Johnson}, {Morton}, {Sinukoff},
  {Crossfield}, \& {Hirsch}}]{FPH17}
{Fulton}, B.~J., {Petigura}, E.~A., {Howard}, A.~W., {et~al.} 2017, Astron. J.,
  154, 109, \dodoi{10.3847/1538-3881/aa80eb}

\bibitem[{{Gardner} {et~al.}(2023){Gardner}, {Mather}, {Abbott}, {Abell},
  {Abernathy}, {Abney}, {Abraham}, {Abraham}, {Abul-Huda}, {Acton}, {Adams},
  {Adams}, {Adler}, {Adriaensen}, {Aguilar}, {Ahmed}, {Ahmed}, {Ahmed},
  {Albat}, {Albert}, {Alberts}, {Aldridge}, {Allen}, {Allen}, {Altenburg},
  {Altunc}, {Alvarez}, {{\'A}lvarez-M{\'a}rquez}, {Alves de Oliveira},
  {Ambrose}, {Anandakrishnan}, {Andersen}, {Anderson}, {Anderson}, {Anderson},
  {Anderson}, {Aprea}, {Archer}, {Arenberg}, {Argyriou}, {Arribas}, {Artigau},
  {Arvai}, {Atcheson}, {Atkinson}, {Averbukh}, {Aymergen}, {Bacinski},
  {Baggett}, {Bagnasco}, {Baker}, {Balzano}, {Banks}, {Baran}, {Barker},
  {Barrett}, {Barringer}, {Barto}, {Bast}, {Baudoz}, {Baum}, {Beatty},
  {Beaulieu}, {Bechtold}, {Beck}, {Beddard}, {Beichman}, {Bellagama}, {Bely},
  {Berger}, {Bergeron}, {Bernier}, {Bertch}, {Beskow}, {Betz}, {Biagetti},
  {Birkmann}, {Bjorklund}, {Blackwood}, {Blazek}, {Blossfeld}, {Bluth},
  {Boccaletti}, {Boegner}, {Bohlin}, {Boia}, {B{\"o}ker}, {Bonaventura},
  {Bond}, {Bosley}, {Boucarut}, {Bouchet}, {Bouwman}, {Bower}, {Bowers},
  {Bowers}, {Boyce}, {Boyer}, {Boyer}, {Boyer}, {Boyer}, {Bradley}, {Brady},
  {Brandl}, {Brannen}, {Breda}, {Bremmer}, {Brennan}, {Bresnahan}, {Bright},
  {Broiles}, {Bromenschenkel}, {Brooks}, {Brooks}, {Brown}, {Brown}, {Brown},
  {Bruce}, {Bryson}, {Bujanda}, {Bullock}, {Bunker}, {Bureo}, {Burt}, {Bush},
  {Bushouse}, {Bussman}, {Cabaud}, {Cale}, {Calhoon}, {Calvani}, {Canipe},
  {Caputo}, {Cara}, {Carey}, {Case}, {Cesari}, {Cetorelli}, {Chance},
  {Chandler}, {Chaney}, {Chapman}, {Charlot}, {Chayer}, {Cheezum}, {Chen},
  {Chen}, {Cherinka}, {Chichester}, {Chilton}, {Chittiraibalan}, {Clampin},
  {Clark}, {Clark}, {Clark}, {Claybrooks}, {Cleveland}, {Cohen}, {Cohen},
  {Col{\'o}n}, {Coleman}, {Colina}, {Comber}, {Comeau}, {Comer}, {Conde Reis},
  {Connolly}, {Conroy}, {Contos}, {Contreras}, {Cook}, {Cooper}, {Cooper},
  {Correia}, {Correnti}, {Cossou}, {Costanza}, {Coulais}, {Cox}, {Coyle},
  {Cracraft}, {Crew}, {Curtis}, {Cusveller}, {Da Costa Maciel}, {Dailey},
  {Daugeron}, {Davidson}, {Davies}, {Davis}, {Davis}, {Day}, {de Chambure}, {de
  Jong}, {De Marchi}, {Dean}, {Decker}, {Delisa}, {Dell}, {Dellagatta},
  {Dembinska}, {Demosthenes}, {Dencheva}, {Deneu}, {DePriest}, {Deschenes},
  {Dethienne}, {Detre}, {Diaz}, {Dicken}, {DiFelice}, {Dillman}, {Disharoon},
  {Dixon}, {Doggett}, {Dominguez}, {Donaldson}, {Doria-Warner}, {Santos},
  {Doty}, {Douglas}, {Doyon}, {Dressler}, {Driggers}, {Driggers}, {Dunn},
  {DuPrie}, {Dupuis}, {Durning}, {Dutta}, {Earl}, {Eccleston}, {Ecobichon},
  {Egami}, {Ehrenwinkler}, {Eisenhamer}, {Eisenhower}, {Eisenstein}, {El
  Hamel}, {Elie}, {Elliott}, {Elliott}, {Engesser}, {Espinoza}, {Etienne},
  {Etxaluze}, {Evans}, {Fabreguettes}, {Falcolini}, {Falini}, {Fatig},
  {Feeney}, {Feinberg}, {Fels}, {Ferdous}, {Ferguson}, {Ferrarese}, {Ferreira},
  {Ferruit}, {Ferry}, {Filippazzo}, {Firre}, {Fix}, {Flagey}, {Flanagan},
  {Fleming}, {Florian}, {Flynn}, {Foiadelli}, {Fontaine}, {Fontanella},
  {Forshay}, {Fortner}, {Fox}, {Framarini}, {Francisco}, {Franck}, {Franx},
  {Franz}, {Friedman}, {Friend}, {Frost}, {Fu}, {Fullerton}, {Gaillard},
  {Galkin}, {Gallagher}, {Galyer}, {Garc{\'\i}a Mar{\'\i}n}, {Gardner},
  {Garland}, {Garrett}, {Gasman}, {G{\'a}sp{\'a}r}, {Gastaud}, {Gaudreau},
  {Gauthier}, {Geers}, {Geithner}, {Gennaro}, {Gerber}, {Gereau}, {Giampaoli},
  {Giardino}, {Gibbons}, {Gilbert}, {Gilman}, {Girard}, {Giuliano}, {Gkountis},
  {Glasse}, {Glassmire}, {Glauser}, {Glazer}, {Goldberg}, {Golimowski},
  {Gonzaga}, {Gordon}, {Gordon}, {Goudfrooij}, {Gough}, {Graham}, {Grau},
  {Green}, {Greene}, {Greene}, {Greenfield}, {Greenhouse}, {Greve}, {Greville},
  {Grimaldi}, {Groe}, {Groebner}, {Grumm}, {Grundy}, {G{\"u}del}, {Guillard},
  {Guldalian}, {Gunn}, {Gurule}, {Gutman}, {Guy}, {Guyot}, {Hack}, {Haderlein},
  {Hagan}, {Hagedorn}, {Hainline}, {Haley}, {Hami}, {Hamilton}, {Hammann},
  {Hammel}, {Hanley}, {Hansen}, {Hardy}, {Harnisch}, {Harr}, {Harris}, {Hart},
  {Hartig}, {Hasan}, {Hashim}, {Hashimoto}, {Haskins}, {Hawkins}, {Hayden},
  {Hayden}, {Healy}, {Hecht}, {Heeg}, {Hejal}, {Helm}, {Hengemihle}, {Henning},
  {Henry}, {Henry}, {Henshaw}, {Hernandez}, {Herrington}, {Heske}, {Hesman},
  {Hickey}, {Hilbert}, {Hines}, {Hinz}, {Hirsch}, {Hitcho}, {Hodapp}, {Hodge},
  {Hoffman}, {Holfeltz}, {Holler}, {Hoppa}, {Horner}, {Howard}, {Howard},
  {Huber}, {Hunkeler}, {Hunter}, {Hunter}, {Hurd}, {Hurst}, {Hutchings},
  {Hylan}, {Ignat}, {Illingworth}, {Irish}, {Isaacs}, {Jackson}, {Jaffe},
  {Jahic}, {Jahromi}, {Jakobsen}, {James}, {James}, {James}, {Jamieson},
  {Jandra}, {Jayawardhana}, {Jedrzejewski}, {Jeffers}, {Jensen}, {Joanne},
  {Johns}, {Johnson}, {Johnson}, {Johnson}, {Johnson}, {Johnson}, {Johnson},
  {Johnstone}, {Jollet}, {Jones}, {Jones}, {Jones}, {Jones}, {Jones}, {Jordan},
  {Jordan}, {Jue}, {Jurkowski}, {Justis}, {Justtanont}, {Kaleida}, {Kalirai},
  {Kalmanson}, {Kaltenegger}, {Kammerer}, {Kan}, {Kanarek}, {Kao}, {Karakla},
  {Karl}, {Kassin}, {Kauffman}, {Kavanagh}, {Kelley}, {Kelly}, {Kendrew},
  {Kennedy}, {Kenny}, {Keski-Kuha}, {Keyes}, {Khan}, {Kidwell}, {Kimble},
  {King}, {King}, {Kinzel}, {Kirk}, {Kirkpatrick}, {Klaassen}, {Klingemann},
  {Klintworth}, {Knapp}, {Knight}, {Knollenberg}, {Knutsen}, {Koehler},
  {Koekemoer}, {Kofler}, {Kontson}, {Kovacs}, {Kozhurina-Platais}, {Krause},
  {Kriss}, {Krist}, {Kristoffersen}, {Krogel}, {Krueger}, {Kulp}, {Kumari},
  {Kwan}, {Kyprianou}, {Labador}, {Labiano}, {Lafreni{\`e}re}, {Lagage},
  {Laidler}, {Laine}, {Laird}, {Lajoie}, {Lallo}, {Lam}, {LaMassa}, {Lambros},
  {Lampenfield}, {Lander}, {Langston}, {Larson}, {Larson}, {LaVerghetta},
  {Law}, {Lawrence}, {Lee}, {Lee}, {Lee}, {Leisenring}, {Leveille}, {Levenson},
  {Levi}, {Levine}, {Lewis}, {Lewis}, {Lewis}, {Libralato}, {Lidon},
  {Liebrecht}, {Lightsey}, {Lilly}, {Lim}, {Lim}, {Ling}, {Link}, {Link},
  {Lipinski}, {Liu}, {Lo}, {Lobmeyer}, {Logue}, {Long}, {Long}, {Long}, {Long},
  {L{\'o}pez-Caniego}, {Lotz}, {Love-Pruitt}, {Lubskiy}, {Luers}, {Luetgens},
  {Luevano}, {Lui}, {Lund}, {Lundquist}, {Lunine}, {L{\"u}tzgendorf}, {Lynch},
  {MacDonald}, {MacDonald}, {Macias}, {Macklis}, {Maghami}, {Maharaja},
  {Maiolino}, {Makrygiannis}, {Malla}, {Malumuth}, {Manjavacas}, {Marini},
  {Marrione}, {Marston}, {Martel}, {Martin}, {Martin}, {Martinez}, {Maschmann},
  {Masci}, {Masetti}, {Maszkiewicz}, {Matthews}, {Matuskey}, {McBrayer},
  {McCarthy}, {McCaughrean}, {McClare}, {McClare}, {McCloskey}, {McClurg},
  {McCoy}, {McElwain}, {McGregor}, {McGuffey}, {McKay}, {McKenzie}, {McLean},
  {McMaster}, {McNeil}, {De Meester}, {Mehalick}, {Meixner}, {Mel{\'e}ndez},
  {Menzel}, {Menzel}, {Merz}, {Mesterharm}, {Meyer}, {Meyett}, {Meza},
  {Midwinter}, {Milam}, {Miller}, {Miller}, {Miskey}, {Misselt}, {Mitchell},
  {Mohan}, {Montoya}, {Moran}, {Morishita}, {Moro-Mart{\'\i}n}, {Morrison},
  {Morrison}, {Morse}, {Moschos}, {Moseley}, {Mosier}, {Mosner}, {Mountain},
  {Muckenthaler}, {Mueller}, {Mueller}, {Muhiem}, {M{\"u}hlmann}, {Mullally},
  {Mullen}, {Munger}, {Murphy}, {Murray}, {Muzerolle}, {Mycroft}, {Myers},
  {Myers}, {Myers}, {Myers}, {Myrick}, {Nagle}, {Nayak}, {Naylor}, {Neff},
  {Nelan}, {Nella}, {Nguyen}, {Nguyen}, {Nickson}, {Nidhiry}, {Niedner},
  {Nieto-Santisteban}, {Nikolov}, {Nishisaka}, {Noriega-Crespo}, {Nota},
  {O'Mara}, {Oboryshko}, {O'Brien}, {Ochs}, {Offenberg}, {Ogle}, {Ohl},
  {Olmsted}, {Osborne}, {O'Shaughnessy}, {{\"O}stlin}, {O'Sullivan}, {Otor},
  {Ottens}, {Ouellette}, {Outlaw}, {Owens}, {Pacifici}, {Page}, {Paranilam},
  {Park}, {Parrish}, {Paschal}, {Patapis}, {Patel}, {Patrick}, {Pattishall},
  {Paul}, {Paul}, {Pauly}, {Pavlovsky}, {Pe{\~n}a-Guerrero}, {Pedder}, {Peek},
  {Pelham}, {Penanen}, {Perriello}, {Perrin}, {Perrine}, {Perrygo}, {Peslier},
  {Petach}, {Peterson}, {Pfarr}, {Pierson}, {Pietraszkiewicz}, {Pilchen},
  {Pipher}, {Pirzkal}, {Pitman}, {Player}, {Plesha}, {Plitzke}, {Pohner},
  {Poletis}, {Pollizzi}, {Polster}, {Pontius}, {Pontoppidan}, {Porges},
  {Potter}, {Prescott}, {Proffitt}, {Pueyo}, {Quispe Neira}, {Radich}, {Rager},
  {Rameau}, {Ramey}, {Ramos Alarcon}, {Rampini}, {Rapp}, {Rashford},
  {Rauscher}, {Ravindranath}, {Rawle}, {Rawlings}, {Ray}, {Regan}, {Rehm},
  {Rehm}, {Reid}, {Reis}, {Renk}, {Reoch}, {Ressler}, {Rest}, {Reynolds},
  {Richon}, {Richon}, {Ridgaway}, {Riedel}, {Rieke}, {Rieke}, {Rifelli},
  {Rigby}, {Riggs}, {Ringel}, {Ritchie}, {Rix}, {Robberto}, {Robinson},
  {Robinson}, {Robinson}, {Rock}, {Rodriguez}, {Rodr{\'\i}guez del Pino},
  {Roellig}, {Rohrbach}, {Roman}, {Romelfanger}, {Romo}, {Rosales}, {Rose},
  {Roteliuk}, {Roth}, {Rothwell}, {Rouzaud}, {Rowe}, {Rowlands}, {Roy},
  {Royer}, {Rui}, {Rumler}, {Rumpl}, {Russ}, {Ryan}, {Ryan}, {Saad}, {Sabata},
  {Sabatino}, {Sabbi}, {Sabelhaus}, {Sabia}, {Sahu}, {Saif}, {Salvignol},
  {Samara-Ratna}, {Samuelson}, {Sanders}, {Sappington}, {Sargent}, {Sauer},
  {Savadkin}, {Sawicki}, {Schappell}, {Scheffer}, {Scheithauer}, {Scherer},
  {Schiff}, {Schlawin}, {Schmeitzky}, {Schmitz}, {Schmude}, {Schneider},
  {Schreiber}, {Schroeven-Deceuninck}, {Schultz}, {Schwab}, {Schwartz},
  {Scoccimarro}, {Scott}, {Scott}, {Seaton}, {Seely}, {Seery}, {Seidleck},
  {Sembach}, {Shanahan}, {Shaughnessy}, {Shaw}, {Shay}, {Sheehan}, {Sheth},
  {Shih}, {Shivaei}, {Siegel}, {Sienkiewicz}, {Simmons}, {Simon}, {Sirianni},
  {Sivaramakrishnan}, {Slade}, {Sloan}, {Slocum}, {Slowinski}, {Smith},
  {Smith}, {Smith}, {Smith}, {Smith}, {Smith}, {Smolik}, {Soderblom}, {Sohn},
  {Sokol}, {Sonneborn}, {Sontag}, {Sooy}, {Soummer}, {Southwood}, {Spain},
  {Sparmo}, {Speer}, {Spencer}, {Sprofera}, {Stallcup}, {Stanley},
  {Stansberry}, {Stark}, {Starr}, {Stassi}, {Steck}, {Steeley}, {Stephens},
  {Stephenson}, {Stewart}, {Stiavelli}, {}, {Strada}, {Straughn}, {Streetman},
  {Strickland}, {Strobele}, {Stuhlinger}, {Stys}, {Such}, {Sukhatme},
  {Sullivan}, {Sullivan}, {Sumner}, {Sun}, {Sunnquist}, {Swade}, {Swam},
  {Swenton}, {Swoish}, {Tam Litten}, {Tamas}, {Tao}, {Taylor}, {Taylor}, {te
  Plate}, {Van Tea}, {Teague}, {Telfer}, {Temim}, {Texter}, {Thatte},
  {Thompson}, {Thompson}, {Thomson}, {Thronson}, {Tierney}, {Tikkanen},
  {Tinnin}, {Tippet}, {Todd}, {Tran}, {Trauger}, {Trejo}, {Vinh Truong},
  {Tsukamoto}, {Tufail}, {Tumlinson}, {Tustain}, {Tyra}, {Ubeda}, {Underwood},
  {Uzzo}, {Vaclavik}, {Valenduc}, {Valenti}, {Van Campen}, {van de Wetering},
  {Van Der Marel}, {van Haarlem}, {Vandenbussche}, {van Dishoeck},
  {Vanterpool}, {Vernoy}, {Vila Costas}, {Volk}, {Voorzaat}, {Voyton}, {Vydra},
  {Waddy}, {Waelkens}, {Wahlgren}, {Walker}, {Wander}, {Warfield}, {Warner},
  {Wasiak}, {Wasiak}, {Wehner}, {Weiler}, {Weilert}, {Weiss}, {Wells}, {Welty},
  {Wheate}, {Wheeler}, {White}, {Whitehouse}, {Whiteleather}, {Whitman},
  {Williams}, {Willmer}, {Willott}, {Willoughby}, {Wilson}, {Wilson}, {Wilson},
  {Windhorst}, {Wislowski}, {Wolfe}, {Wolfe}, {Wolff}, {Wondel}, {Woo},
  {Woods}, {Worden}, {Workman}, {Wright}, {Wu}, {Wu}, {Wun}, {Wymer},
  {Yadetie}, {Yan}, {Yang}, {Yates}, {Yeager}, {Yerger}, {Young}, {Young},
  {Yu}, {Yu}, {Zak}, {Zeidler}, {Zepp}, {Zhou}, {Zincke}, {Zonak}, \&
  {Zondag}}]{GMA23}
{Gardner}, J.~P., {Mather}, J.~C., {Abbott}, R., {et~al.} 2023, Publ. Astron.
  Soc. Pac., 135, 068001, \dodoi{10.1088/1538-3873/acd1b5}

\bibitem[{{Ginzburg} {et~al.}(2018){Ginzburg}, {Schlichting}, \& {Sari}}]{GSS}
{Ginzburg}, S., {Schlichting}, H.~E., \& {Sari}, R. 2018, Mon. Not. R. Astron.
  Soc., 476, 759, \dodoi{10.1093/mnras/sty290}

\bibitem[{{Greene} {et~al.}(2023){Greene}, {Bell}, {Ducrot}, {Dyrek}, {Lagage},
  \& {Fortney}}]{GBD23}
{Greene}, T.~P., {Bell}, T.~J., {Ducrot}, E., {et~al.} 2023, Nature, 618, 39,
  \dodoi{10.1038/s41586-023-05951-7}

\bibitem[{{Gronoff} {et~al.}(2020){Gronoff}, {Arras}, {Baraka}, {Bell},
  {Cessateur}, {Cohen}, {Curry}, {Drake}, {Elrod}, {Erwin}, {Garcia-Sage},
  {Garraffo}, {Glocer}, {Heavens}, {Lovato}, {Maggiolo}, {Parkinson}, {Simon
  Wedlund}, {Weimer}, \& {Moore}}]{GAB20}
{Gronoff}, G., {Arras}, P., {Baraka}, S., {et~al.} 2020, J. Geophys. Res. Space
  Phys., 125, e27639, \dodoi{10.1029/2019JA027639}

\bibitem[{{Gunell} {et~al.}(2018){Gunell}, {Maggiolo}, {Nilsson}, {Stenberg
  Wieser}, {Slapak}, {Lindkvist}, {Hamrin}, \& {De Keyser}}]{GMN18}
{Gunell}, H., {Maggiolo}, R., {Nilsson}, H., {et~al.} 2018, Astron. Astrophys.,
  614, L3, \dodoi{10.1051/0004-6361/201832934}

\bibitem[{{Gupta} \& {Schlichting}(2019)}]{GS19}
{Gupta}, A., \& {Schlichting}, H.~E. 2019, Mon. Not. R. Astron. Soc., 487, 24,
  \dodoi{10.1093/mnras/stz1230}

\bibitem[{{Gupta} \& {Schlichting}(2020)}]{GS20}
---. 2020, Mon. Not. R. Astron. Soc., 493, 792, \dodoi{10.1093/mnras/staa315}

\bibitem[{{Kane}(2022)}]{SRK22}
{Kane}, S.~R. 2022, Nat. Astron., 6, 420, \dodoi{10.1038/s41550-022-01626-x}

\bibitem[{{Kane} {et~al.}(2019){Kane}, {Arney}, {Crisp}, {Domagal-Goldman},
  {Glaze}, {Goldblatt}, {Grinspoon}, {Head}, {Lenardic}, {Unterborn}, {Way}, \&
  {Zahnle}}]{KAC19}
{Kane}, S.~R., {Arney}, G., {Crisp}, D., {et~al.} 2019, J. Geophys. Res.
  Planets, 124, 2015, \dodoi{10.1029/2019JE005939}

\bibitem[{{Kasting} {et~al.}(1993){Kasting}, {Whitmire}, \& {Reynolds}}]{KWR93}
{Kasting}, J.~F., {Whitmire}, D.~P., \& {Reynolds}, R.~T. 1993, Icarus, 101,
  108, \dodoi{10.1006/icar.1993.1010}

\bibitem[{{Kopparapu} {et~al.}(2014){Kopparapu}, {Ramirez}, {SchottelKotte},
  {Kasting}, {Domagal-Goldman}, \& {Eymet}}]{KRS14}
{Kopparapu}, R.~K., {Ramirez}, R.~M., {SchottelKotte}, J., {et~al.} 2014,
  Astrophys. J. Lett., 787, L29, \dodoi{10.1088/2041-8205/787/2/L29}

\bibitem[{{Kopparapu} {et~al.}(2013){Kopparapu}, {Ramirez}, {Kasting}, {Eymet},
  {Robinson}, {Mahadevan}, {Terrien}, {Domagal-Goldman}, {Meadows}, \&
  {Deshpande}}]{KRK13}
{Kopparapu}, R.~K., {Ramirez}, R., {Kasting}, J.~F., {et~al.} 2013, Astrophys.
  J., 765, 131, \dodoi{10.1088/0004-637X/765/2/131}

\bibitem[{{Krenn} {et~al.}(2021){Krenn}, {Fossati}, {Kubyshkina}, \&
  {Lammer}}]{KFKL}
{Krenn}, A.~F., {Fossati}, L., {Kubyshkina}, D., \& {Lammer}, H. 2021, Astron.
  Astrophys., 650, A94, \dodoi{10.1051/0004-6361/202140437}

\bibitem[{{Kubyshkina} {et~al.}(2018){Kubyshkina}, {Fossati}, {Erkaev},
  {Cubillos}, {Johnstone}, {Kislyakova}, {Lammer}, {Lendl}, \& {Odert}}]{KFE18}
{Kubyshkina}, D., {Fossati}, L., {Erkaev}, N.~V., {et~al.} 2018, Astrophys. J.
  Lett., 866, L18, \dodoi{10.3847/2041-8213/aae586}

\bibitem[{{Lammer}(2013)}]{HL13}
{Lammer}, H. 2013, {Origin and Evolution of Planetary Atmospheres} (Berlin:
  Springer), \dodoi{10.1007/978-3-642-32087-3}

\bibitem[{{Lammer} {et~al.}(2008){Lammer}, {Kasting}, {Chassefi{\`e}re},
  {Johnson}, {Kulikov}, \& {Tian}}]{LKC08}
{Lammer}, H., {Kasting}, J.~F., {Chassefi{\`e}re}, E., {et~al.} 2008, Space
  Sci. Rev., 139, 399, \dodoi{10.1007/s11214-008-9413-5}

\bibitem[{{Lamp{\'o}n} {et~al.}(2021){Lamp{\'o}n}, {L{\'o}pez-Puertas},
  {Czesla}, {S{\'a}nchez-L{\'o}pez}, {Lara}, {Salz}, {Sanz-Forcada},
  {Molaverdikhani}, {Quirrenbach}, {Pall{\'e}}, {Caballero}, {Henning},
  {Nortmann}, {Amado}, {Montes}, {Reiners}, \& {Ribas}}]{LLC21}
{Lamp{\'o}n}, M., {L{\'o}pez-Puertas}, M., {Czesla}, S., {et~al.} 2021, Astron.
  Astrophys., 648, L7, \dodoi{10.1051/0004-6361/202140423}

\bibitem[{{Laneuville} {et~al.}(2020){Laneuville}, {Dong}, {O’Rourke}, \&
  {Schneider}}]{LDO20}
{Laneuville}, M., {Dong}, C., {O’Rourke}, J.~G., \& {Schneider}, A.~C. 2020,
  in Planetary Diversity: Rocky planet processes and their observational
  signatures, ed. E.~J. {Tasker}, C.~{Unterborn}, M.~{Laneuville}, Y.~{Fujii},
  S.~J. {Desch}, \& H.~E. {Hartnett} (Bristol: IOP Publishing), 3.1--3.47

\bibitem[{{Lichtenegger} {et~al.}(2022){Lichtenegger}, {Dyadechkin}, {Scherf},
  {Lammer}, {Adam}, {Kallio}, {Amerstorfer}, \& {Jarvinen}}]{LDS}
{Lichtenegger}, H.~I.~M., {Dyadechkin}, S., {Scherf}, M., {et~al.} 2022,
  Icarus, 382, 115009, \dodoi{10.1016/j.icarus.2022.115009}

\bibitem[{{Lingam}(2019)}]{LM19}
{Lingam}, M. 2019, Astrophys. J. Lett., 874, L28,
  \dodoi{10.3847/2041-8213/ab12eb}

\bibitem[{{Lingam} \& {Loeb}(2019)}]{LL19}
{Lingam}, M., \& {Loeb}, A. 2019, Rev. Mod. Phys., 91, 021002,
  \dodoi{10.1103/RevModPhys.91.021002}

\bibitem[{{Lingam} \& {Loeb}(2021)}]{ML21}
---. 2021, {Life in the Cosmos: From Biosignatures to Technosignatures}
  (Cambridge: Harvard University Press)

\bibitem[{{Lopez} \& {Fortney}(2014)}]{LF14}
{Lopez}, E.~D., \& {Fortney}, J.~J. 2014, Astrophys. J., 792, 1,
  \dodoi{10.1088/0004-637X/792/1/1}

\bibitem[{{Lustig-Yaeger} {et~al.}(2023){Lustig-Yaeger}, {Fu}, {May},
  {Ceballos}, {Moran}, {Peacock}, {Stevenson}, {Kirk}, {L{\'o}pez-Morales},
  {MacDonald}, {Mayorga}, {Sing}, {Sotzen}, {Valenti}, {Redai}, {Alam},
  {Batalha}, {Bennett}, {Gonzalez-Quiles}, {Kruse}, {Lothringer},
  {Rustamkulov}, \& {Wakeford}}]{LFM23}
{Lustig-Yaeger}, J., {Fu}, G., {May}, E.~M., {et~al.} 2023, Nat. Astron., 7,
  1317, \dodoi{10.1038/s41550-023-02064-z}

\bibitem[{{Madhusudhan}(2019)}]{NM19}
{Madhusudhan}, N. 2019, Annu. Rev. Astron. Astrophys., 57, 617,
  \dodoi{10.1146/annurev-astro-081817-051846}

\bibitem[{{Martinez} {et~al.}(2019){Martinez}, {Cunha}, {Ghezzi}, \&
  {Smith}}]{MCG19}
{Martinez}, C.~F., {Cunha}, K., {Ghezzi}, L., \& {Smith}, V.~V. 2019,
  Astrophys. J., 875, 29, \dodoi{10.3847/1538-4357/ab0d93}

\bibitem[{{McDonald} {et~al.}(2019){McDonald}, {Kreidberg}, \& {Lopez}}]{MKL19}
{McDonald}, G.~D., {Kreidberg}, L., \& {Lopez}, E. 2019, Astrophys. J., 876,
  22, \dodoi{10.3847/1538-4357/ab1095}

\bibitem[{{McIntyre} {et~al.}(2019){McIntyre}, {Lineweaver}, \&
  {Ireland}}]{MLI19}
{McIntyre}, S. R.~N., {Lineweaver}, C.~H., \& {Ireland}, M.~J. 2019, Mon. Not.
  R. Astron. Soc., 485, 3999, \dodoi{10.1093/mnras/stz667}

\bibitem[{{Mendillo} {et~al.}(2018){Mendillo}, {Withers}, \& {Dalba}}]{MWD18}
{Mendillo}, M., {Withers}, P., \& {Dalba}, P.~A. 2018, Nat. Astron., 2, 287,
  \dodoi{10.1038/s41550-017-0375-y}

\bibitem[{{Mordasini}(2020)}]{M20}
{Mordasini}, C. 2020, Astron. Astrophys., 638, A52,
  \dodoi{10.1051/0004-6361/201935541}

\bibitem[{{Ostberg} {et~al.}(2023){Ostberg}, {Kane}, {Li}, {Schwieterman},
  {Hill}, {Bott}, {Dalba}, {Fetherolf}, {Head}, \& {Unterborn}}]{OKL23}
{Ostberg}, C., {Kane}, S.~R., {Li}, Z., {et~al.} 2023, arXiv e-prints,
  arXiv:2302.03055, \dodoi{10.48550/arXiv.2302.03055}

\bibitem[{{Otegi} {et~al.}(2020){Otegi}, {Bouchy}, \& {Helled}}]{OBH20}
{Otegi}, J.~F., {Bouchy}, F., \& {Helled}, R. 2020, Astron. Astrophys., 634,
  A43, \dodoi{10.1051/0004-6361/201936482}

\bibitem[{{Owen}(2019)}]{JO19}
{Owen}, J.~E. 2019, Annu. Rev. Earth Planet. Sci., 47, 67,
  \dodoi{10.1146/annurev-earth-053018-060246}

\bibitem[{{Owen} \& {Alvarez}(2016)}]{OA16}
{Owen}, J.~E., \& {Alvarez}, M.~A. 2016, Astrophys. J., 816, 34,
  \dodoi{10.3847/0004-637X/816/1/34}

\bibitem[{{Owen} \& {Jackson}(2012)}]{OJ12}
{Owen}, J.~E., \& {Jackson}, A.~P. 2012, Mon. Not. R. Astron. Soc., 425, 2931,
  \dodoi{10.1111/j.1365-2966.2012.21481.x}

\bibitem[{{Owen} \& {Schlichting}(2023)}]{OS23}
{Owen}, J.~E., \& {Schlichting}, H.~E. 2023, Mon. Not. R. Astron. Soc.,
  \dodoi{10.1093/mnras/stad3972}

\bibitem[{{Owen} \& {Wu}(2017)}]{OW17}
{Owen}, J.~E., \& {Wu}, Y. 2017, Astrophys. J., 847, 29,
  \dodoi{10.3847/1538-4357/aa890a}

\bibitem[{{Perryman}(2018)}]{MP18}
{Perryman}, M. 2018, {The Exoplanet Handbook}, 2nd edn. (Cambridge: Cambridge
  University Press)

\bibitem[{{Persson} {et~al.}(2020){Persson}, {Futaana}, {Ramstad}, {Masunaga},
  {Nilsson}, {Hamrin}, {Fedorov}, \& {Barabash}}]{PFR20}
{Persson}, M., {Futaana}, Y., {Ramstad}, R., {et~al.} 2020, J. Geophys. Res.
  Planets, 125, e06336, \dodoi{10.1029/2019JE006336}

\bibitem[{{Plotnykov} \& {Valencia}(2020)}]{PV20}
{Plotnykov}, M., \& {Valencia}, D. 2020, Mon. Not. R. Astron. Soc., 499, 932,
  \dodoi{10.1093/mnras/staa2615}

\bibitem[{{Ramstad} \& {Barabash}(2021)}]{RB21}
{Ramstad}, R., \& {Barabash}, S. 2021, Space Sci. Rev., 217, 36,
  \dodoi{10.1007/s11214-021-00791-1}

\bibitem[{{Rogers} {et~al.}(2021){Rogers}, {Gupta}, {Owen}, \&
  {Schlichting}}]{RGO21}
{Rogers}, J.~G., {Gupta}, A., {Owen}, J.~E., \& {Schlichting}, H.~E. 2021, Mon.
  Not. R. Astron. Soc., 508, 5886, \dodoi{10.1093/mnras/stab2897}

\bibitem[{{Salz} {et~al.}(2016){Salz}, {Schneider}, {Czesla}, \&
  {Schmitt}}]{SSC16}
{Salz}, M., {Schneider}, P.~C., {Czesla}, S., \& {Schmitt}, J.~H.~M.~M. 2016,
  Astron. Astrophys., 585, L2, \dodoi{10.1051/0004-6361/201527042}

\bibitem[{{Seager} \& {Deming}(2010)}]{SD10}
{Seager}, S., \& {Deming}, D. 2010, Annu. Rev. Astron. Astrophys., 48, 631,
  \dodoi{10.1146/annurev-astro-081309-130837}

\bibitem[{{Shields} {et~al.}(2016){Shields}, {Ballard}, \& {Johnson}}]{SBJ16}
{Shields}, A.~L., {Ballard}, S., \& {Johnson}, J.~A. 2016, Phys. Rep., 663, 1,
  \dodoi{10.1016/j.physrep.2016.10.003}

\bibitem[{{Tian}(2015)}]{FT15}
{Tian}, F. 2015, Annu. Rev. Earth Planet. Sci., 43, 459,
  \dodoi{10.1146/annurev-earth-060313-054834}

\bibitem[{{Tinetti} {et~al.}(2018){Tinetti}, {Drossart}, {Eccleston},
  {Hartogh}, {Heske}, {Leconte}, {Micela}, {Ollivier}, {Pilbratt}, {Puig},
  {Turrini}, {Vandenbussche}, {Wolkenberg}, {Beaulieu}, {Buchave}, {Ferus},
  {Griffin}, {Guedel}, {Justtanont}, {Lagage}, {Machado}, {Malaguti}, {Min},
  {N{\o}rgaard-Nielsen}, {Rataj}, {Ray}, {Ribas}, {Swain}, {Szabo}, {Werner},
  {Barstow}, {Burleigh}, {Cho}, {Coud{\'e} du Foresto}, {Coustenis}, {Decin},
  {Encrenaz}, {Galand}, {Gillon}, {Helled}, {Morales}, {Garc{\'\i}a Mu{\~n}oz},
  {Moneti}, {Pagano}, {Pascale}, {Piccioni}, {Pinfield}, {Sarkar}, {Selsis},
  {Tennyson}, {Triaud}, {Venot}, {Waldmann}, {Waltham}, {Wright}, {Amiaux},
  {Augu{\`e}res}, {Berth{\'e}}, {Bezawada}, {Bishop}, {Bowles}, {Coffey},
  {Colom{\'e}}, {Crook}, {Crouzet}, {Da Peppo}, {Sanz}, {Focardi}, {Frericks},
  {Hunt}, {Kohley}, {Middleton}, {Morgante}, {Ottensamer}, {Pace}, {Pearson},
  {Stamper}, {Symonds}, {Rengel}, {Renotte}, {Ade}, {Affer}, {Alard}, {Allard},
  {Altieri}, {Andr{\'e}}, {Arena}, {Argyriou}, {Aylward}, {Baccani}, {Bakos},
  {Banaszkiewicz}, {Barlow}, {Batista}, {Bellucci}, {Benatti}, {Bernardi},
  {B{\'e}zard}, {Blecka}, {Bolmont}, {Bonfond}, {Bonito}, {Bonomo}, {Brucato},
  {Brun}, {Bryson}, {Bujwan}, {Casewell}, {Charnay}, {Pestellini}, {Chen},
  {Ciaravella}, {Claudi}, {Cl{\'e}dassou}, {Damasso}, {Damiano}, {Danielski},
  {Deroo}, {Di Giorgio}, {Dominik}, {Doublier}, {Doyle}, {Doyon}, {Drummond},
  {Duong}, {Eales}, {Edwards}, {Farina}, {Flaccomio}, {Fletcher}, {Forget},
  {Fossey}, {Fr{\"a}nz}, {Fujii}, {Garc{\'\i}a-Piquer}, {Gear}, {Geoffray},
  {G{\'e}rard}, {Gesa}, {Gomez}, {Graczyk}, {Griffith}, {Grodent}, {Guarcello},
  {Gustin}, {Hamano}, {Hargrave}, {Hello}, {Heng}, {Herrero}, {Hornstrup},
  {Hubert}, {Ida}, {Ikoma}, {Iro}, {Irwin}, {Jarchow}, {Jaubert}, {Jones},
  {Julien}, {Kameda}, {Kerschbaum}, {Kervella}, {Koskinen}, {Krijger}, {Krupp},
  {Lafarga}, {Landini}, {Lellouch}, {Leto}, {Luntzer}, {Rank-L{\"u}ftinger},
  {Maggio}, {Maldonado}, {Maillard}, {Mall}, {Marquette}, {Mathis}, {Maxted},
  {Matsuo}, {Medvedev}, {Miguel}, {Minier}, {Morello}, {Mura}, {Narita},
  {Nascimbeni}, {Nguyen Tong}, {Noce}, {Oliva}, {Palle}, {Palmer}, {Pancrazzi},
  {Papageorgiou}, {Parmentier}, {Perger}, {Petralia}, {Pezzuto},
  {Pierrehumbert}, {Pillitteri}, {Piotto}, {Pisano}, {Prisinzano}, {Radioti},
  {R{\'e}ess}, {Rezac}, {Rocchetto}, {Rosich}, {Sanna}, {Santerne}, {Savini},
  {Scandariato}, {Sicardy}, {Sierra}, {Sindoni}, {Skup}, {Snellen}, {Sobiecki},
  {Soret}, {Sozzetti}, {Stiepen}, {Strugarek}, {Taylor}, {Taylor}, {Terenzi},
  {Tessenyi}, {Tsiaras}, {Tucker}, {Valencia}, {Vasisht}, {Vazan}, {Vilardell},
  {Vinatier}, {Viti}, {Waters}, {Wawer}, {Wawrzaszek}, {Whitworth}, {Yung},
  {Yurchenko}, {Zapatero Osorio}, {Zellem}, {Zingales}, \& {Zwart}}]{TDE}
{Tinetti}, G., {Drossart}, P., {Eccleston}, P., {et~al.} 2018, Exp. Astron.,
  46, 135, \dodoi{10.1007/s10686-018-9598-x}

\bibitem[{{TRAPPIST-1 JWST Community Initiative} {et~al.}(2023){TRAPPIST-1 JWST
  Community Initiative}, {de Wit}, {Doyon}, {Rackham}, {Lim}, {Ducrot},
  {Kreidberg}, {Benneke}, {Ribas}, {Berardo}, {Niraula}, {Iyer}, {Shapiro},
  {Kostogryz}, {Witzke}, {Gillon}, {Agol}, {Meadows}, {Burgasser}, {Owen},
  {Fortney}, {Selsis}, {Bello-Arufe}, {Bolmont}, {Cowan}, {Dong}, {Drake},
  {Garcia}, {Greene}, {Haworth}, {Hu}, {Kane}, {Kervella}, {Koll},
  {Krissansen-Totton}, {Lagage}, {Lichtenberg}, {Lustig-Yaeger}, {Lingam},
  {Turbet}, {Seager}, {Barkaoui}, {Bell}, {Burdanov}, {Cadieux}, {Charnay},
  {Cloutier}, {Cook}, {Correia}, {Dang}, {Daylan}, {Delrez}, {Edwards},
  {Fauchez}, {Flagg}, {Fraschetti}, {Haqq-Misra}, {Huang}, {Iro},
  {Jayawardhana}, {Jehin}, {Jin}, {Kite}, {Kitzmann}, {Kral}, {Lafreni{\`e}re},
  {Libert}, {Liu}, {Mohanty}, {Morris}, {Murray}, {Piaulet}, {Pozuelos},
  {Radica}, {Ranjan}, {Rathcke}, {Roy}, {Schwieterman}, {Turner}, {Triaud}, \&
  {Way}}]{TJCI2023}
{TRAPPIST-1 JWST Community Initiative}, {de Wit}, J., {Doyon}, R., {et~al.}
  2023, arXiv e-prints, arXiv:2310.15895, \dodoi{10.48550/arXiv.2310.15895}

\bibitem[{{Valencia} {et~al.}(2006){Valencia}, {O'Connell}, \&
  {Sasselov}}]{VOS06}
{Valencia}, D., {O'Connell}, R.~J., \& {Sasselov}, D. 2006, Icarus, 181, 545,
  \dodoi{10.1016/j.icarus.2005.11.021}

\bibitem[{{Van Eylen} {et~al.}(2018){Van Eylen}, {Agentoft}, {Lundkvist},
  {Kjeldsen}, {Owen}, {Fulton}, {Petigura}, \& {Snellen}}]{VAL18}
{Van Eylen}, V., {Agentoft}, C., {Lundkvist}, M.~S., {et~al.} 2018, Mon. Not.
  R. Astron. Soc., 479, 4786, \dodoi{10.1093/mnras/sty1783}

\bibitem[{{Watson} {et~al.}(1981){Watson}, {Donahue}, \& {Walker}}]{WDW81}
{Watson}, A.~J., {Donahue}, T.~M., \& {Walker}, J.~C.~G. 1981, Icarus, 48, 150,
  \dodoi{10.1016/0019-1035(81)90101-9}

\bibitem[{{Winn} \& {Fabrycky}(2015)}]{WF15}
{Winn}, J.~N., \& {Fabrycky}, D.~C. 2015, Annu. Rev. Astron. Astrophys., 53,
  409, \dodoi{10.1146/annurev-astro-082214-122246}

\bibitem[{{Wood} {et~al.}(2021){Wood}, {M{\"u}ller}, {Redfield}, {Konow},
  {Vannier}, {Linsky}, {Youngblood}, {Vidotto}, {Jardine},
  {Alvarado-G{\'o}mez}, \& {Drake}}]{WMR21}
{Wood}, B.~E., {M{\"u}ller}, H.-R., {Redfield}, S., {et~al.} 2021, Astrophys.
  J., 915, 37, \dodoi{10.3847/1538-4357/abfda5}

\bibitem[{{Wordsworth} \& {Kreidberg}(2022)}]{WK22}
{Wordsworth}, R., \& {Kreidberg}, L. 2022, Annu. Rev. Astron. Astrophys., 60,
  159, \dodoi{10.1146/annurev-astro-052920-125632}

\bibitem[{{Zahnle} \& {Catling}(2017)}]{ZC17}
{Zahnle}, K.~J., \& {Catling}, D.~C. 2017, Astrophys. J., 843, 122,
  \dodoi{10.3847/1538-4357/aa7846}

\bibitem[{{Zeng} {et~al.}(2016){Zeng}, {Sasselov}, \& {Jacobsen}}]{ZSJ16}
{Zeng}, L., {Sasselov}, D.~D., \& {Jacobsen}, S.~B. 2016, Astrophys. J., 819,
  127, \dodoi{10.3847/0004-637X/819/2/127}

\bibitem[{{Zeng} {et~al.}(2019){Zeng}, {Jacobsen}, {Sasselov}, {Petaev},
  {Vanderburg}, {Lopez-Morales}, {Perez-Mercader}, {Mattsson}, {Li}, {Heising},
  {Bonomo}, {Damasso}, {Berger}, {Cao}, {Levi}, \& {Wordsworth}}]{ZJS19}
{Zeng}, L., {Jacobsen}, S.~B., {Sasselov}, D.~D., {et~al.} 2019, Proc. Natl.
  Acad. Sci., 116, 9723, \dodoi{10.1073/pnas.1812905116}

\bibitem[{{Zhang}(2020)}]{XZ20}
{Zhang}, X. 2020, Res. Astron. Astrophys., 20, 099,
  \dodoi{10.1088/1674-4527/20/7/99}

\bibitem[{{Zhu} \& {Dong}(2021)}]{ZD21}
{Zhu}, W., \& {Dong}, S. 2021, Annu. Rev. Astron. Astrophys., 59, 291,
  \dodoi{10.1146/annurev-astro-112420-020055}

\bibitem[{{Zieba} {et~al.}(2023){Zieba}, {Kreidberg}, {Ducrot}, {Gillon},
  {Morley}, {Schaefer}, {Tamburo}, {Koll}, {Lyu}, {Acu{\~n}a}, {Agol}, {Iyer},
  {Hu}, {Lincowski}, {Meadows}, {Selsis}, {Bolmont}, {Mandell}, \&
  {Suissa}}]{ZKD23}
{Zieba}, S., {Kreidberg}, L., {Ducrot}, E., {et~al.} 2023, \nat, 620, 746,
  \dodoi{10.1038/s41586-023-06232-z}

\end{thebibliography}

\end{document}